\documentclass[preprint,pra,showpacs,nofootinbib]{revtex4}
\usepackage{mathrsfs}
\usepackage{graphicx}
\usepackage{epsfig}
\usepackage{dcolumn}
\usepackage{bm}
\usepackage{amsmath,amssymb,amsthm}
\usepackage[colorlinks=true,linkcolor=blue]{hyperref}
\usepackage{subfigure}
\usepackage{booktabs}
\usepackage[mathscr]{euscript}
\usepackage{ulem}

\newcommand{\bea}{\begin{eqnarray}}
\newcommand{\eea}{\end{eqnarray}}
\newcommand{\beq}{\begin{equation}}
\newcommand{\eeq}{\end{equation}}

\def\/{\over}

\begin{document}
\title{Coherent structure of\\ Alice-Bob modified Korteweg de-Vries  Equation}
\author{Congcong Li$^{1}$, S. Y. Lou $^{1,2}$\footnote{Corresponding author: lou@nbu.edu.cn}and Man Jia $^{1}$}

\affiliation{
$^{1}$\footnotesize{Center for Nonlinear Science and Department of Physics, Ningbo University, Ningbo, 315211, China}\\
$^2$\footnotesize Shanghai Key Laboratory of Trustworthy Computing, East China Normal University, Shanghai 200062, China }

\begin{abstract}
To describe two-place events, Alice-Bob systems have been established by means of the shifted parity and delayed time reversal in Ref. \cite{AB}. In this paper, we mainly study exact solutions of the integrable Alice-Bob modified Korteweg de-Vries (AB-mKdV) system. The general Nth Darboux transformation for the AB-mKdV equation are constructed. By using the Darboux transformation, some types of shifted parity and time reversal symmetry breaking solutions including one-soliton, two-soliton and rogue wave solutions are explicitly obtained. In addition to the similar solutions of the mKdV equation (group invariant solutions), there are abundant new localized structures for the AB-mKdV systems.  
\end{abstract}

\maketitle

\section{Introduction}
In 2013, Ablowitz and Musslimani \cite{1} proposed a new integrable nonlocal nonlinear Schr\"odinger (NLS) equation
\begin{equation}
iA_{t}+A_{xx}\pm A^2B=0, \ \ \  B=\hat{f}A=\hat{P}\hat{C}A=A^*(-x,t), \label{Eq1}
\end{equation}
where the operators $\hat{P}$ and $\hat{C}$ are the usual parity and charge conjugation. In literature, the nonlocal nonlinear Schr\"odinger equation \eqref{Eq1} is also called parity-time reversal (PT) symmetric. PT symmetry plays an important role in the quantum physics \cite{2} and many other areas of physics, such as quantum chromodynamics \cite{3}, electric circuits \cite{4}, optics \cite{5,6} and Bose-Einstein condensates \cite{7}, etc.

It is well known that there are various correlated and/or entangled events that may be happened in different times and places. To describe two-place physical problems, Alice-Bob (AB) systems \cite{AB} are proposed by using the AB-BA equivalence principle and the shifted parity ($\hat{P}_{s}$), delayed time reversal ($\hat{T}_{d}$) and charge conjugate ($\hat{C}$) symmetries. If one event (A, Alice event) is correlated/entangled to another (B, Bob event), we denote the correlated relation as $B=\hat{f}A$ for suitable $\hat{f}$ operators. Usually, the event $A=A(x,\ t)$ happened at $\{x,\ t\}$ and event $B=B(x',\ t')$ happened at $\{x',\ t'\}=\hat{f}\{x,\ t\}$. In fact, $\{x',\ t'\}$ is usually far away from $\{x,\ t\}$. Hence, the intrinsic two-place models or Alice-Bob systems are nonlocal. In addition to the nonlocal nonlinear Schr\"odinger equation \eqref{Eq1}, there are many other types of two-place nonlocal models, such as the nonlocal KdV systems \cite{9}, the nonlocal modified KdV systems \cite{10,11}, the discrete nonlocal NLS systems \cite{12}, the coupled nonlocal NLS systems \cite{13} and the nonlocal Davey-Stewartson systems \cite{14,15,16}, etc.

In \cite{AB}, one of us (Lou) proposed a series of integrable AB systems including the AB-KdV systems, AB-mKdV systems, AB-KP systems, AB-sine Gordon systems, AB-NLS systems and AB-Toda systems. Furthermore, by using the $\hat{P}_{s},\ \hat{T}_{d}$ and $C$ symmetries, their $\hat{P}_{s},\ \hat{T}_{d}$ and $\hat{C}$ invariant muti-soliton solutions are obtained in elegant forms.
 In addition, Lou established a most general AB-KdV equation and presented its $\hat{P}_{s},\ \hat{T}_{d}$ and $\hat{C}$ invariant Painlev\'e II reduction and soliton-cnoidal periodic wave interaction solutions for the AB-KdV system \cite{17}. However, to find $\hat{P}_{s},\ \hat{T}_{d}$ and $\hat{C}$ symmetry breaking solutions is much more difficult. 
 
In this paper, influenced by the idea of Lou in \cite{17}, we try to investigate $\hat{P}_{s},\ \hat{T}_{d}$ and $\hat{C}$ symmetry breaking solutions for a special AB-mKdV equation which has been also proposed in \cite{AB}. On the one hand, we will show that it can be derived from the third order AKNS system. On the other hand, we will construct its Nth Darboux transformation and give its one-soliton solutions and two-soliton solutions through Darboux transformation. These explicit solutions possess some new properties that are different from the ones for the mKdV equation.

\section{AB-mKdV systems and their common shifted parity and time reversal invariant solutions}

The most general AB-mKdV system 
may have the form 
\begin{equation}
\left\{\begin{array}{l}
K(A,\ B)=0,\\
B=\hat{f}A=\pm \hat{P_{s}}\hat{T_{d}}\hat{C}^a A=\pm \hat{C}^a A(-x+x_{0},-t+t_{0}),\ a=0,\ \mbox{\rm or}\ 1,
\end{array}\right.\label{KAB}
\end{equation}
where $x_0$ and $t_0$ are arbitrary constants and $K(A,\ B)$ is an arbitrary functional of $A$ and $B$ with the condition 
\begin{equation}
K(u,\ u)=u_t+u_{xxx}+6u^2u_x=0.\label{Kuu} 
\end{equation} 
In fact here $\hat{f}=\pm \hat{P}_s\hat{T}_d\hat{C}^a$ is a discrete symmetry of the mKdV equation \eqref{Kuu} with $\hat{f}^2=1$. 

A concrete differential polynomial form of \eqref{KAB}
reads
\begin{equation}
\left\{\begin{array}{l}
A_t+b_1A_{xxx}+b_2B_{xxx}+(a_1A^2+a_2AB+a_3B^2)A_x+(a_4A^2+a_5AB+a_6 B^2)B_x
=0,\\
B=\hat{f}A=\pm \hat{C}^a A(-x+x_{0},-t+t_{0}),\ a=0,\ \mbox{\rm or}\ 1, \ 
\end{array}\right.\label{KABpoly}
\end{equation}
where $a_1,\ a_2,\ a_3,\ a_4,\ a_5$ and $b_1$ are arbitrary constants while $b_2=1-b_1$ and $a_6=6-a_1-a_2-a_3-a_4-a_5$. The AB-mKdV system \eqref{KABpoly} can be considered as a special reduction of the coupled mKdV equation which can be derived from the two layer fluid dynamic systems \cite{CmKdV}. 

A more special form of the AB-mKdV system,
\begin{eqnarray}
\begin{aligned}
&A_{t}+A_{xxx}+6ABA_{x}=0,\\
&B=\hat{f}A=\pm \hat{P_{s}}\hat{T_{d}}\hat{C}^aA=\pm \hat{C}^a A(-x+x_{0},-t+t_{0}), \ a=0,\ \mbox{\rm or}\ 1,
\end{aligned}\label{ABmKdV}
\end{eqnarray}
can also be considered as a special reduction of 
 the third order AKNS system \cite{AB},
\begin{equation}
AKNS3\equiv
\left\{\begin{array}{ll}
A_{t}+A_{xxx}+6ABA_{x}=0, \\
B_{t}+B_{xxx}+6ABB_{x}=0.
\end{array}
\right.
\end{equation}

Because of the property \eqref{Kuu}, all the shifted parity and delayed time reversal invariant (for simplicity, $\hat{f}$-invariant) solutions of the general AB-mKdV system \eqref{KAB} possess the same form of the usual mKdV equation \eqref{Kuu}. Thus, to find $\hat{f}$-invariant solutions of the AB-mKdV systems is equivalent to select out the $\hat{f}$-invariant solutions from known solutions (which are usually $\hat{f}$ symmetry breaking) of the usual mKdV equation. 

It is fortunate that the multiple soliton solutions of the mKdV equation can be reconstructed as \cite{AB}, 
\begin{eqnarray}
u_{soliton}=
\pm 2\frac{\partial}{\partial x} \tan^{-1}\frac{\sum_{\nu_e}K_{\nu}\sinh\left(\sum_{j=1}^N \nu_j\eta_j\right)}{\sum_{\nu_o}K_{\nu}\cosh\left(\sum_{j=1}^N \nu_j\eta_j\right)}, \label{Nsoliton}
\end{eqnarray}
where 
the summation of $\nu_o$ should be done for all non-dual odd  permutations of $\nu_i=1,\ -1, \ i=1,\ 2,\ \ldots,\ N$ with odd number of $\nu_i=1$, the summation of $\nu_e$ should be done for all non-dual even permutations of $\nu_i=1,\ -1, \ i=1,\ 2,\ \ldots,\ N$ with even number of 
$\nu_i=1$, 
\begin{equation}
K_\nu \equiv \prod_{i>j}(k_i-\nu_i\nu_jk_j),
\end{equation}
 and $\eta_j$ being defined as 
\begin{eqnarray}
\eta_j=k_j\left(x-\frac12x_0\right)-k_j^3\left(t-\frac12t_0\right)+\eta_{0j},\label{eta}
\end{eqnarray} 
$k_j,\ j=1,\ 2,\ \ldots,\ N$ and 
$\eta_{0j},\ j=1,\ 2,\ \ldots,\ N$
are arbitrary constants.  

It is clear that if some of $\eta_{0j}$ are nonzero, then the multiple solutions \eqref{Nsoliton} is $\hat{f}$ symmetry breaking. The arbitrariness of $\eta_{0j}$ is introduced by the space-time translation invariants. However, all the AB-mKdV systems are space-time translation symmetry breaking. Thus, $\eta_{0j}$ should be fixed. From the expressions \eqref{Nsoliton} and \eqref{eta}, it is straightforward to find that 
\begin{equation}
A=\left. u_{soliton}\right|_{\eta_{0j}=0,\ j=1,\ \ldots,\ N}
\end{equation}
is just the $\hat{f}$-invariant $N$-soliton solution of all the real AB-mKdV systems \eqref{KAB}.  

Because all the $\hat{f}$-invariant solutions of the AB-mKdV systems only constitute a subset of the solutions of the mKdV equation, it is more interest to find $\hat{f}$-symmetry breaking solutions of the AB-mKdV systems. In order to find some nontrivial $\hat{f}$-symmetry breaking solutions, we restrict to study the Darboux transformations of the special AB-mKdV equation \eqref{ABmKdV} with $a=0$ and the lower negative sign, i.e., $\hat{B}=-A(-x+x_0,\ -t+t_0)$.

\section{Darboux transformation of the AB-mKdV System}

The Darboux transformation method can be traced back to the way of thinking in the study of the linear problem of Darboux. It is an effective method to obtain exact solutions for integrable nonlinear systems \cite{19}. In this section, we will give the Darboux transformation for the AB-mKdV system \eqref{ABmKdV}. First, we start with the following Lax pair of the AB-mKdV system \eqref{ABmKdV}:
\begin{equation}
\varphi_{x}=U\varphi=
\left(\begin{array}{cc}
-i\lambda & A \\
B & i\lambda \\
\end{array}
\right)\varphi,\ \ \ 
\varphi_{t}=V\varphi=\left( \begin{array}{cc}
 \alpha & \beta \\
\gamma & -\alpha \\
\end{array} 
\right)\varphi, \label{Lax}
\end{equation}
where $\varphi=(\varphi_{1}(x,t),\varphi_{2}(x,t))^{T}$, $\lambda$ is the spectral parameter and $\alpha$, $\beta$ and $\gamma$ are given by ($i=\sqrt{-1}$)
\begin{eqnarray}
\begin{aligned}
&\alpha=-4i\lambda^{3}
-2iAB\lambda+AB_{x}-BA_{x},\\
&\beta=4A\lambda^{2}
+{2}iA_{x}\lambda+2A^{2}B-A_{xx},\\
&\gamma=4B\lambda^{2}
-2iB_{x}\lambda+2AB^{2}-B_{xx}.
\end{aligned}
\end{eqnarray}
The compatibility condition of equation \eqref{Lax},
 $$U_{t}-V_{x}+[U,V]=0$$
  results in \eqref{ABmKdV}.

Secondly, imitating the procedure of Darboux transformation for general integrable mKdV equation \cite{20,21}, 
we will construct the Darboux transformation of the AB-mKdV equation \eqref{ABmKdV}. Taking the gauge transformation,
\begin{equation}
\varphi^{[1]}=T^{[1]}\varphi,
\end{equation}
the spectral problem \eqref{Lax} turns into
\begin{eqnarray}
\begin{aligned}
&\varphi_{x}^{[1]}=(T_{x}^{[1]}+T^{[1]}U)(T^{[1]})^{-1}\varphi^{[1]}=U^{[1]}\varphi^{[1]},\\ &\varphi_{t}^{[1]}=(T_{t}^{[1]}+T^{[1]}V)(T^{[1]})^{-1}\varphi^{[1]}=V^{[1]}\varphi^{[1]}.
\end{aligned}
\end{eqnarray}
Letting 
\begin{equation}
T^{[1]}=\lambda I+S^{[1]},
\end{equation}
with $S^{[1]}=(s_{ij}^{[1]})_{2\times2}$, $s_{ij}^{[1]}(i,j=1,2)$ are functions of $x$ and $t$, $I$ being the identity matrix. After that, we  get the relationship between the new potentials $\{A^{[1]},\ B^{[1]}\}$ and the old ones $\{A,\ B\}$,
\begin{eqnarray}
\begin{aligned}
&A^{[1]}=A+2is_{12}^{[1]},\\
&B^{[1]}=B-2is_{21}^{[1]}.
\end{aligned}
\end{eqnarray}
From the correlation relation 
$B=\hat{f}A=- A(-x+x_0,\ -t+t_0)$, we obtain the following constraint:
\begin{equation}
s_{12}^{[1]}(-x+x_0,-t+t_0)=s_{21}^{[1]}(x,t).\label{BfA}
\end{equation}

The eigenfunctions corresponding to the seed solution are 
$$f(\lambda_{j})=
(f_1(\lambda_{j}),\ f_2(\lambda_{j}))^{T},\  g(\lambda_{j})
=(g_1(\lambda_{j}),\ g_2(\lambda_{j}))^{T}$$ and the eigenvalues are $\lambda=\lambda_{j}(j=1,2)$ in \eqref{Lax}. Then we get 
\begin{eqnarray}
\begin{aligned}
&\lambda_{j}+s_{11}^{[1]}+\alpha_{j}s_{12}^{[1]}=0,\\
&s_{21}^{[1]}+\alpha_{j}(\lambda_{j}+s_{22}^{[1]})=0,\\
&\alpha_{j}=\frac{f_2(\lambda_{j})+\gamma_{j}g_2(\lambda_{j})}{f_1(\lambda_{j})+\gamma_{j}g_1(\lambda_{j})},
\end{aligned}
\end{eqnarray}
with $\gamma_{j}(j=1,2)$ being arbitrary constants.

Thus, the matrix $T^{[1]}$ can be written as
\begin{equation}
T^{[1]}=\left(
          \begin{array}{cc}
            \lambda & 0 \\
            0 & \lambda \\
          \end{array}
        \right)+\frac{1}{\alpha_2-\alpha_1}\left(
                                             \begin{array}{cc}
                                               \lambda_2\alpha_1-\lambda_1\alpha_2 & \lambda_1-\lambda_2 \\
                                               \alpha_1\alpha_2(\lambda_2-\lambda_1) & \lambda_1\alpha_1-\lambda_2\alpha_2 \\
                                             \end{array}
                                           \right).
\end{equation}
Finally, we construct the n-fold Darboux transformation for the AB-mKdV system (2) to find the $P_{s}T_{d}$ symmetry breaking soliton solutions.
\begin{equation}
\varphi^{[n]}=T_{n}(\lambda)\varphi,  \ \ \  T_{n}(\lambda)=T^{[n]}(\lambda)T^{[n-1]}(\lambda)\cdots T^{[k]}(\lambda)\cdots T^{[1]}(\lambda),
\end{equation}
with
\begin{equation}
T^{[k]}(\lambda)=\lambda I+S^{[k]}=\lambda I+\frac{1}{\alpha_{2k}-\alpha_{2k-1}}\left(
                                                                                  \begin{array}{cc}
                                                                                    \lambda_{2k}\alpha_{2k-1}-\lambda_{2k-1}\alpha_{2k} & \lambda_{2k-1}-\lambda_{2k} \\
                                                                                    \alpha_{2k-1}\alpha_{2k}(\lambda_{2k}-\lambda_{2k-1}) & \lambda_{2k-1}\alpha_{2k-1}-\lambda_{2k}\alpha_{2k} \\
                                                                                  \end{array}
                                                                                \right),
\end{equation}
where
\begin{eqnarray}
\begin{aligned}
&\alpha_{j}=\frac{f_{2}^{[k-1]}(\lambda_{j})+\gamma_{j}g_{2}^{[k-1]}(\lambda_{j})}{f_{1}^{[k-1]}(\lambda_{j})+\gamma_{j}g_{1}^{[k-1]}(\lambda_{j})},\ \ \ (j=2k-1,2k,\ \ \ k=1,2,\cdots ,n)\\
&f^{[k]}(\lambda)=\left(
                    \begin{array}{c}
                      f_{1}^{[k]}(\lambda) \\
                      f_{2}^{[k]}(\lambda) \\
                    \end{array}
                  \right)=T^{[k]}(\lambda)f^{[k-1]}(\lambda_{1},\lambda_{2},\cdots ,\lambda_{2k-1},\lambda_{2k}),\\
&g^{[k]}(\lambda)=\left(
                    \begin{array}{c}
                      g_{1}^{[k]}(\lambda) \\
                      g_{2}^{[k]}(\lambda) \\
                    \end{array}
                  \right)=T^{[k]}(\lambda)g^{[k-1]}(\lambda_{1},\lambda_{2},\cdots ,\lambda_{2k-1},\lambda_{2k}),
\end{aligned}
\end{eqnarray}
and the matrix $S^{[k]}$ meets the following constraint condition
\begin{equation}
s_{12}^{[k]}(-x+x_{0},-t+t_{0})=s_{21}^{[k]}(x,t) \ \ \ (k=1,2,\cdots ,n).\label{sk}
\end{equation}
The new solution $A^{[n]}(x,t)$ and old one $A(x,t)$ should satisfy
\begin{equation}
A^{[n]}=A+2i\sum_{k=1}^{n} s_{12}^{[k]}.
\end{equation}
\section{Soliton solutions of the AB-mKdV System}

In this section, we will describe how to obtain the exact solutions of the AB-mKdV system \eqref{ABmKdV} in detail, including one-soliton solutions and two-soliton solutions with the help of the Darboux transformation.

\subsection{One-soliton solutions from zero seed}

The well known solution with exponential form of the AB-mKdV equation \eqref{ABmKdV} can be written in the following form,
\begin{equation}
A=\rho e^{\kappa((x-\frac{x_{0}}{2})-(\kappa^2+6\rho^2)(t-\frac{t_{0}}{2}))},
\end{equation}
where $\kappa$ and $\rho$ are two complex parameters.

First of all, we choose zero seed solution $A=0$. By solving the spectral equation corresponding to zero seed, we get 
\begin{equation}
f(x,t;\lambda)=\left(
                 \begin{array}{c}
                   e^{-i\lambda(x+4\lambda^2t)} \\
                   0 \\
                 \end{array}
               \right),\ \ \ g(x,t;\lambda)=\left(
                 \begin{array}{c}
                   0 \\
                   e^{i\lambda(x+4\lambda^2t)} \\
                 \end{array}
               \right).
\end{equation}
Hence, we obtain
\begin{eqnarray}
\begin{aligned}
&\alpha_{j}=\gamma_{j}e^{2i\lambda_{j}(x+4\lambda_{j}^2t)}=\gamma_{j}e^{\xi_{j}},\ \ \ j=1,2\\
&s_{12}(x,t)=\frac{\lambda_{1}-\lambda_{2}}{\gamma_{2}e^{\xi_{2}}-\gamma_{1}e^{\xi_{1}}},\ \ \ s_{21}(x,t)=\frac{(\lambda_{2}-\lambda_{1})\gamma_{1}\gamma_{2}e^{\xi_{1}+\xi_{2}}}{\gamma_{2}e^{\xi_{2}}-\gamma_{1}e^{\xi_{1}}}.
\end{aligned}
\end{eqnarray}
The constraint condition \eqref{BfA} results in
\begin{equation}
\gamma_{1}^2=e^{-\xi_{1}(x_{0},t_{0})},\ \ \ \gamma_{2}^2=e^{-\xi_{2}(x_{0},t_{0})},
\end{equation}
where $\xi_{j}(x_0,t_0)=2i\lambda_{j}(x_0+4\lambda_{j}^2t_0),(j=1,2)$.\\
We consider the case of $\gamma_{1}=-e^{-\frac{1}{2}\xi_{1}(x_{0},t_{0})}$,$\gamma_{2}=e^{-\frac{1}{2}\xi_{2}(x_{0},t_{0})}$, thus, 
\begin{equation}
A^{[1]}=\frac{2i(\lambda_{1}-\lambda_{2})}{e^{\xi_{1}-\frac{1}{2}\xi_{1}(x_{0},t_{0})}+e^{\xi_{2}-\frac{1}{2}\xi_{2}(x_{0},t_{0})}} \label{1s}
\end{equation}
Note that
\begin{equation}
|e^{\xi_{1}}+e^{\xi_{2}}|^2=2e^{\xi_{1R}+\xi_{2R}}(\cosh(\xi_{1R}-\xi_{2R})+\cos(\xi_{1I}-\xi_{2I})),
\end{equation}
where $\lambda_{j}=\mu_{j}+i\nu_{j}$, $\mu_{j}$, $\nu_{j}$ $\in R\ (j=1,\ 2)$, and $\xi_{jR}$ and $\xi_{jI}$ are real and imaginary parts of $\xi_j$ respectively, 
\begin{eqnarray}
\begin{aligned}
&\xi_{jR}=\Re(\xi_{j})=-2\nu_{j}x+8\nu_{j}(\nu_{j}^2-3\mu_{j}^2)t,\\
&\xi_{jI}=\Im(\xi_{j})=2\mu_{j}x+8\mu_{j}(\mu_{j}^2-3\nu_{j}^2)t.\\
\end{aligned}
\end{eqnarray}
In order to guarantee the solution does not include any singular points, $\lambda_{j}$(j=1,2) should satisfy
\begin{eqnarray}
\begin{aligned}
& 2\mu_{1}\nu_{1}+\mu_{2}\nu_{1}+\mu_{1}\nu_{2}+2\mu_{2}\nu_{2}=0,\\
& (\nu_{1}-\nu_{2})^6\ \ +\left[ \nu_{1}(\nu_{1}^2-3\mu_{1}^2)-\nu_{2}(\nu_{2}^2-3\mu_{2}^2)\right]^2\neq0.
\end{aligned}
\end{eqnarray}
Letting $\mu_{1}=\mu_{2}$ and $\nu_{1}=-\nu_{2}$, then $\xi_{1R}+\xi_{2R}=0$ holds for all $(x,t)\in R^2$, and meets the above conditions. Therefore, we gain a typical soliton,
\begin{eqnarray}
\begin{aligned}
&A^{[1]}=-2i\nu_{1}e^{-i\mu_{1}\zeta_{1}}\mbox {\rm sech}(\nu_{1}\zeta_{2}),\label{A1}\\
&\zeta_{1}=2\left(x-\frac{x_{0}}{2}\right)+8\big(\mu_{1}^2-3\nu_{1}^2\big)\left(t-\frac{t_{0}}{2}\right),\\
&\zeta_{2}=2\left(x-\frac{x_{0}}{2}\right)-8\big(\nu_{1}^2-3\mu_{1}^2\big)\left(t-\frac{t_{0}}{2}\right).
\end{aligned}
\end{eqnarray}
Thus, the soliton propagates to the right when $\nu_{1}^2-3\mu_{1}^2>0$ while the soliton spreads to the left if $\nu_{1}^2-3\mu_{1}^2<0$. As $\nu_{1}^2-3\mu_{1}^2=0$, the soliton is stationary. Fig.1 shows this situation.

\input epsf
     \begin{figure}
     \epsfxsize=7cm\epsfysize=5cm\epsfbox{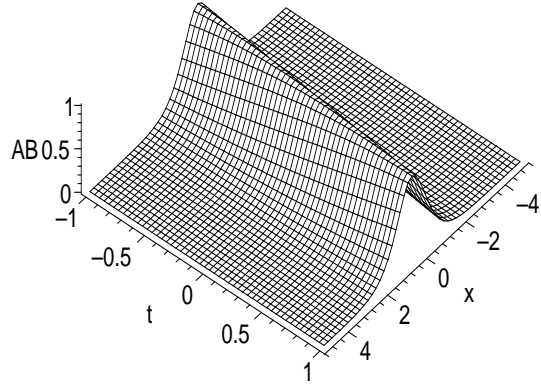}
\caption{Plot of the single soliton \eqref{A1} with $\mu_{1}=0.2$, $\nu_{1}=0.5$, $x_{0}=2$ and $t_{0}=2$ for the quantity $AB$.}
\end{figure}

Setting $\nu_{2}=0$ and $2\mu_{1}+\mu_{2}=0$. This results in $\xi_{2R}=0$ for all $(x,t)\in R^2$. Now the solution gives 
\begin{eqnarray}
\begin{aligned}
&A^{[1]}=\frac{(-2\nu_{1}+6i\mu_{1})e^{2i\mu_{1}\zeta_{1}}}{1+e^{(3i\mu_{1}-\nu_{1})\zeta_{2}}},\\
&\zeta_{1}=2(x-\frac{x_{0}}{2})+32\mu_{1}^2(t-\frac{t_{0}}{2}),\\
&\zeta_{2}=2(x-\frac{x_{0}}{2})-8(\nu_{1}^2-3\mu_{1}^2)(t-\frac{t_{0}}{2}).
\end{aligned} \label{34}
\end{eqnarray}
From this solution, we know that if $\nu_{1}>0$, the complexiton spreads like kink, if $\nu_{1}<0$, it presents the antikink-shape. Similarly, as $\nu_{1}^2-3\mu_{1}^2>0$, the wave travels to the right, as $\nu_{1}^2-3\mu_{1}^2<0$, it propagates to the left, while $\nu_{1}^2-3\mu_{1}^2=0$, it is stationary. Fig.2 describes the situation.

\input epsf
     \begin{figure}
\epsfxsize=7cm\epsfysize=5cm\epsfbox{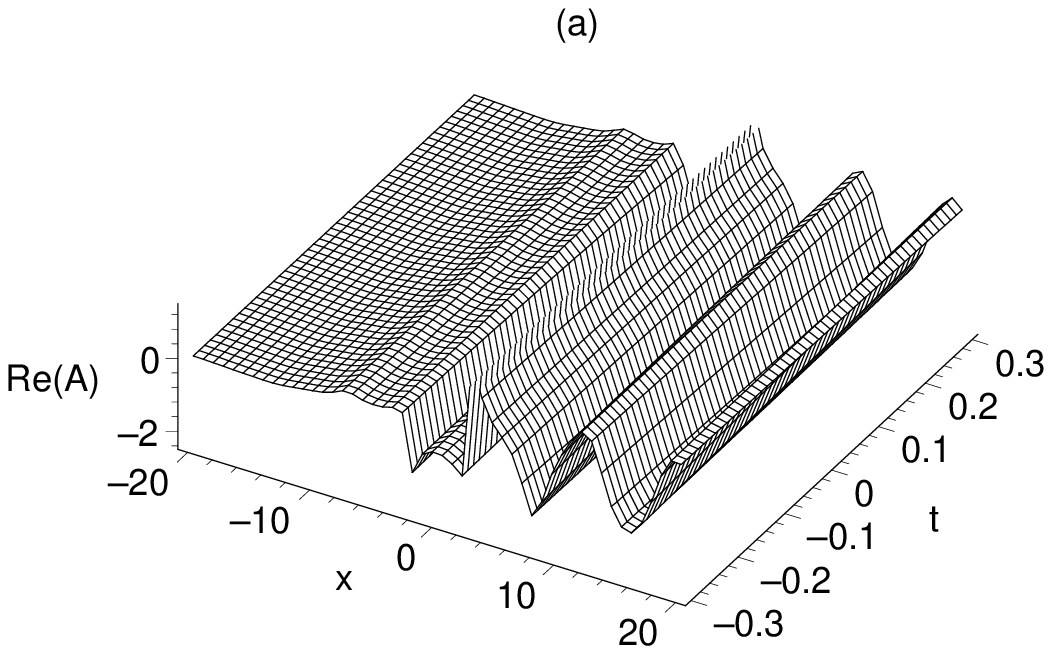}
\epsfxsize=7cm\epsfysize=5cm\epsfbox{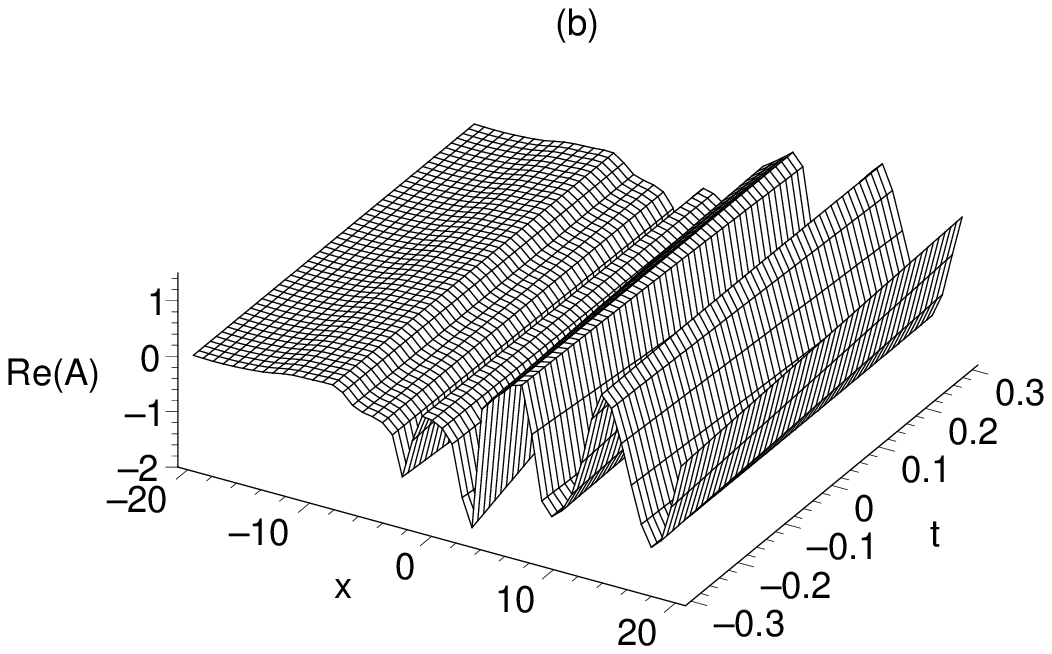}
\epsfxsize=7cm\epsfysize=5cm\epsfbox{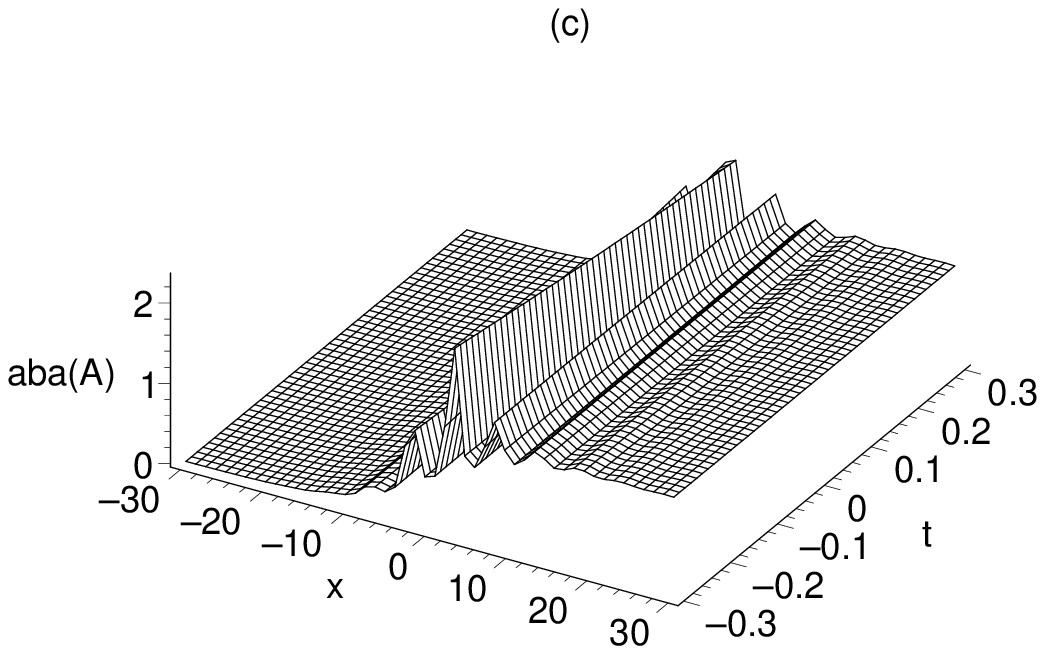}
\caption{Display of the structure of the kink shape complexiton solution \eqref{34} with $\mu_{1}=0.1$, $\nu_{1}=0.1$ and $x_{0}=t_{0}=0$ for (a) the real part $Re(A)=\Re(A)$, (b) the imaginary part $Im(A)=\Im(A)$ and (c) the amplitude $M(A)$.}
\end{figure}

\subsection{One-soliton solutions from nonzero seed.}

To find the one soliton solution from nonzero seed of the AB-mKdV equation \eqref{ABmKdV} with $B=-A(-x+x_0,\ t=-t+t_0)$, we suppose that the nonzero seed $A(x,t)=\rho e^{\kappa((x-\frac{x_{0}}{2})-(\kappa^2+6\rho^2)(t-\frac{t_{0}}{2}))}$, $\rho\neq0$. As the eigenvalue $\lambda_{1}=\frac{i}{2}(\kappa+2\rho)$, we obtain eigenfunctions 
\begin{eqnarray}
\begin{aligned}
&\varphi_{1}(x,t)=\left\{c_{1}+(c_{1}+c_{2})\left[\rho (x-\frac{x_{0}}{2})-\delta_{+}(t-\frac{t_{0}}{2})\right]\right\}e^{\frac{\kappa}{2}\left[(x-\frac{x_{0}}{2})-(\kappa^2+6\rho^2)(t-\frac{t_{0}}{2})\right]},\\
&\varphi_{2}(x,t)=\left\{c_{2}-(c_{1}+c_{2})\left[\rho (x-\frac{x_{0}}{2})-\delta_{+}(t-\frac{t_{0}}{2})\right]\right\}e^{-\frac{\kappa}{2}\left[(x-\frac{x_{0}}{2})-(\kappa^2+6\rho^2)(t-\frac{t_{0}}{2})\right]}.
\end{aligned}
\end{eqnarray}
where $\delta_{+}=3\kappa^2\rho+6\kappa\rho^2+6\rho^3$, $c_{1}$, $c_{2}$ are arbitrary constants.\\
By taking $c_{1}=1$, $c_{2}=0$, which leads to 
\begin{equation}
\alpha_{1}=e^{-\kappa\left[(x-\frac{x_{0}}{2})-(\kappa^2+6\rho^2)(t-\frac{t_{0}}{2})\right]}\left\{-1+\frac{\gamma_{1}+1}{1+(\gamma_{1}+1)\left[\rho (x-\frac{x_{0}}{2})-\delta_{+}(t-\frac{t_{0}}{2})\right]}\right\},
\end{equation}
In the case of $\lambda_{2}=\frac{i}{2}(\kappa-2\rho)$, the eigenfunctions have the following expression,
\begin{eqnarray}
\begin{aligned}
&\varphi_{3}(x,t)=\left\{c_{3}+(c_{4}-c_{3})\left[\rho (x-\frac{x_{0}}{2})-\delta_{-}(t-\frac{t_{0}}{2})\right]\right\} e^{\frac{\kappa}{2}\left[(x-\frac{x_{0}}{2})-(\kappa^2+6\rho^2)(t-\frac{t_{0}}{2})\right]},\\
&\varphi_{4}(x,t)=\left\{c_{4}+(c_{4}-c_{3})\left[\rho (x-\frac{x_{0}}{2})-\delta_{-}(t-\frac{t_{0}}{2})\right]\right\}e^{-\frac{\kappa}{2}\left[(x-\frac{x_{0}}{2})-(\kappa^2+6\rho^2)(t-\frac{t_{0}}{2})\right]}.
\end{aligned}
\end{eqnarray}
where $\delta_{-}=3\kappa^2\rho-6\kappa\rho^2+6\rho^3$, $c_{3}$, $c_{4}$ are arbitrary constants.\\
By setting $c_{3}=0$, $c_{4}=1$, which yields to
\begin{equation}
\alpha_{2}=e^{-\kappa\left[(x-\frac{x_{0}}{2})-(\kappa^2+6\rho^2)(t-\frac{t_{0}}{2})\right]}\left\{1+\frac{\gamma_{2}-1}{1+(\gamma_{2}-1)\left[\rho (x-\frac{x_{0}}{2})-\delta_{-}(t-\frac{t_{0}}{2})\right]}\right\}.
\end{equation}
By solving the constraint condition \eqref{BfA}, we get $\gamma_{1}^2=1$, $\gamma_{2}^2=1$. Then we set $\gamma_{1}=1$, $\gamma_{2}=-1$. Thus the general expression of one-soliton solutions is
\begin{eqnarray}
\begin{aligned}
&A^{[1]}=\rho e^{\kappa\left[(x-\frac{x_{0}}{2})-(\kappa^2+6\rho^2)(t-\frac{t_{0}}{2})\right]}\left(-1+\frac{4-48\xi_{1}}{1+4\rho^2\xi_{2}^2-144\xi_{1}^2}\right), \\
&\xi_{1}=\kappa \rho^2\left(t-\frac{t_{0}}{2}\right),\\
&\xi_{2}=3(\kappa^2+2\rho^2)\left(t-\frac{t_{0}}{2}\right)-\left(x-\frac{x_{0}}{2}\right)
\end{aligned}\label{RW}
\end{eqnarray}
Usually, the expression \eqref{RW} displays some kinds of rogue wave structure for the quantity $AB$. 

The solution \eqref{RW} is $\hat{P}_s\hat{T}_d\hat{C}$ invariant for imaginary $\kappa$ and real $\rho$. However, \eqref{RW} is not $-\hat{P}_s\hat{T}_d$ invariant for any selections of parameters. 

  For pure imaginary $\kappa$ and real $\rho$, setting $\kappa=i\kappa_{I}$ and $\rho=\rho_{R}$, equation \eqref{RW} is changed to
\begin{eqnarray}
\begin{aligned}
&A^{[1]}=\rho_{R} e^{i\kappa_{I}\left[(x-\frac{x_{0}}{2})+(\kappa_{I}^2-6\rho_{R}^2)(t-\frac{t_{0}}{2})\right]}\left(-1+\frac{4-48i\xi_{1I}}{1+4\rho_{R}^2\xi_{2R}^2+144\xi_{1I}^2}\right),\label{Rogue}\\
&\xi_{1I}=\kappa_{I} \rho_{R}^2\left(t-\frac{t_{0}}{2}\right),\\
&\xi_{2R}=3\left(\kappa_{I}^2+2\rho_{R}^2\right)\left(t-\frac{t_{0}}{2}\right)+\left(x-\frac{x_{0}}{2}\right)
\end{aligned}
\end{eqnarray}
In Fig.3, we describe such a rogue-wave with $\rho=1$ and $\kappa=i$. In this case, we see that an eye-shaped form which has a hump and two valleys. 

\input epsf
     \begin{figure}
     \epsfxsize=7cm\epsfysize=5cm\epsfbox{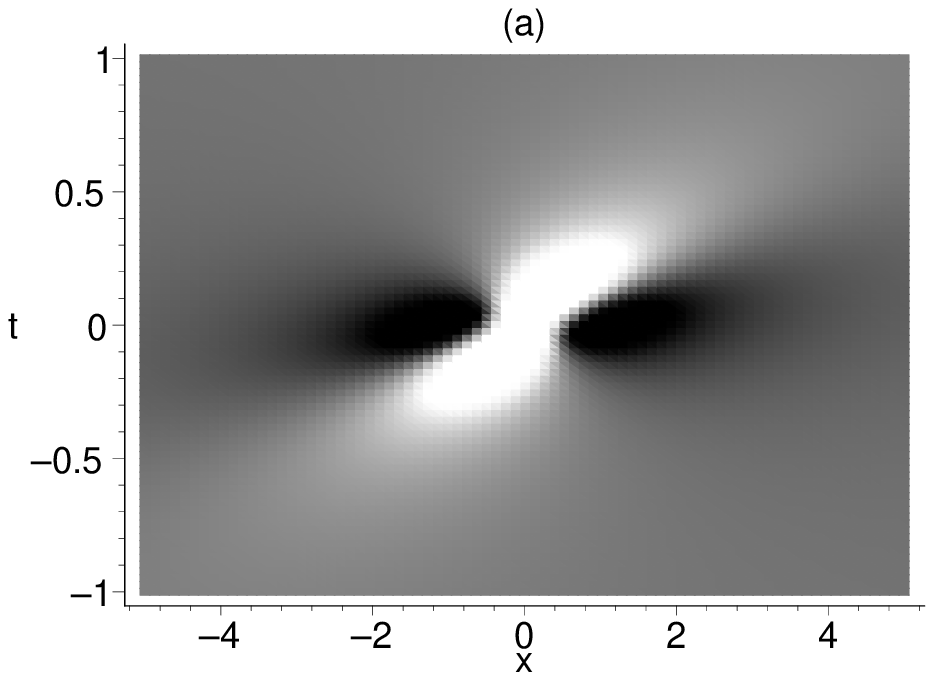}
     \epsfxsize=7cm\epsfysize=5cm\epsfbox{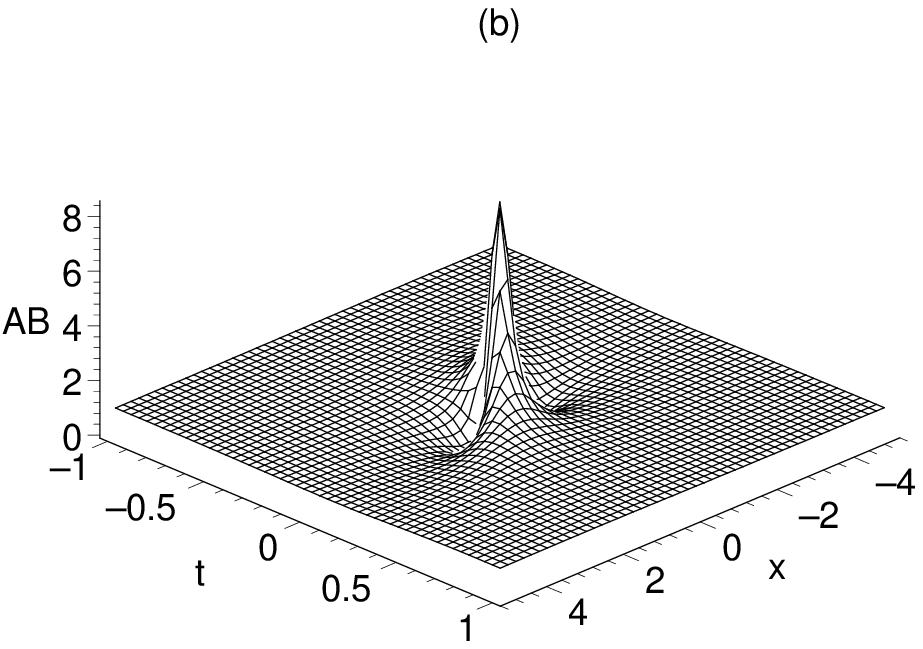}
     \caption{Rogue-wave solution \eqref{Rogue} with $\rho=1$, $\kappa=i$, $x_{0}=0$ and $t_{0}=0$. Its density plot and shape are described in (a) and (b) respectively for the quantity $AB$.}
\end{figure}

For general selections of parameters, the structure of the rogue wave \eqref{Rogue} may be complicate and even singular. In Fig. 4, a different analytic rogue wave is displayed for its real part, imaginary part and amplitude with the parameter selections,
\begin{equation}
\kappa=i+0.1,\ \rho=1-0.1i,\ x_0=t_0=0. \label{ccR}
\end{equation} 

\input epsf
     \begin{figure}
     \epsfxsize=7cm\epsfysize=5cm\epsfbox{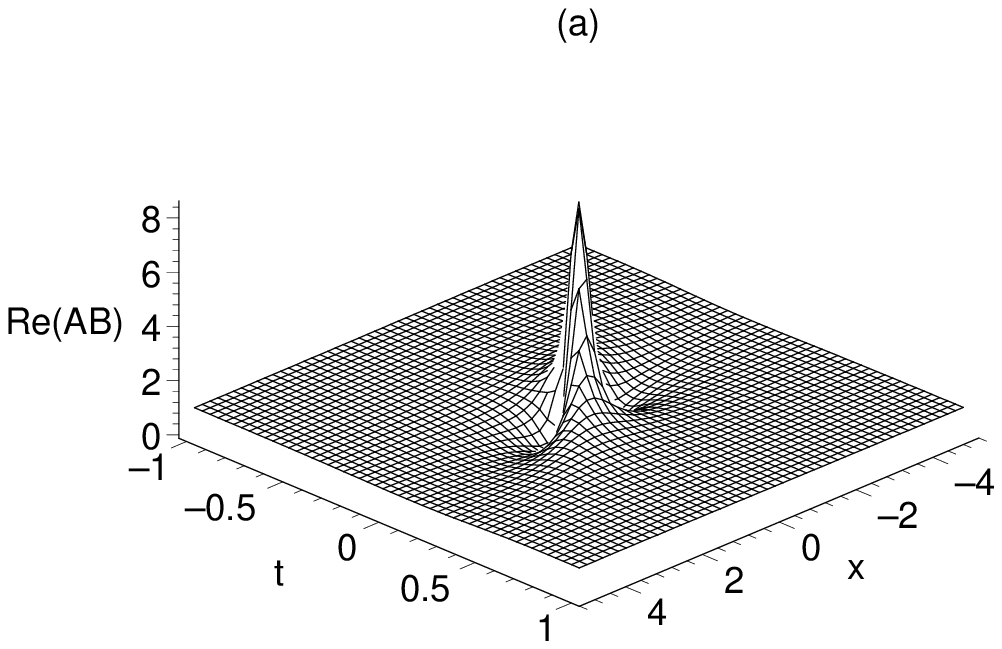}
\epsfxsize=7cm\epsfysize=5cm\epsfbox{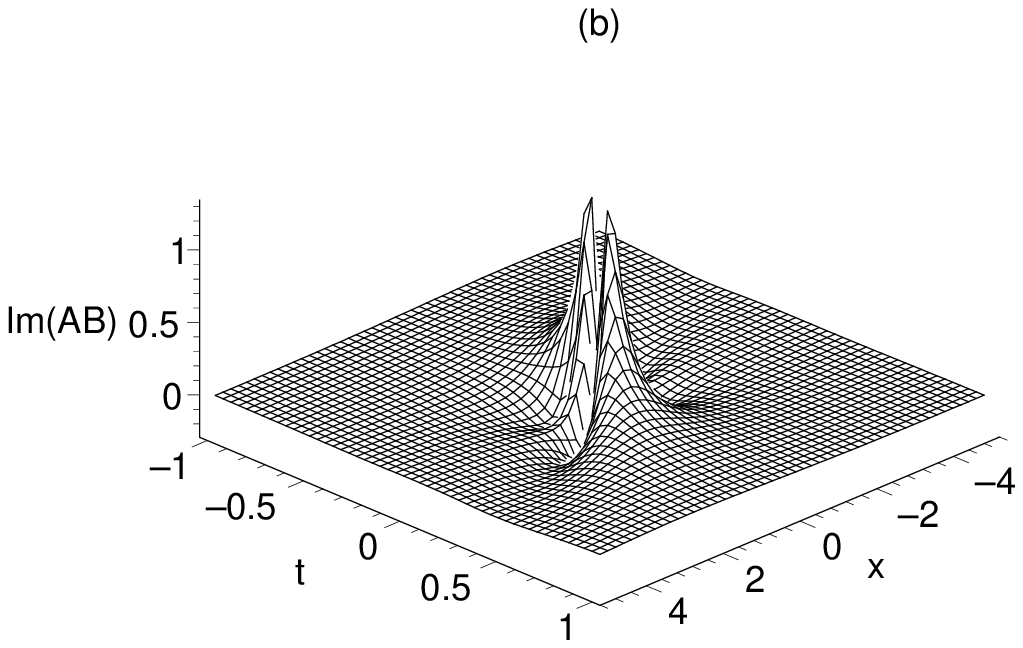}
\epsfxsize=7cm\epsfysize=5cm\epsfbox{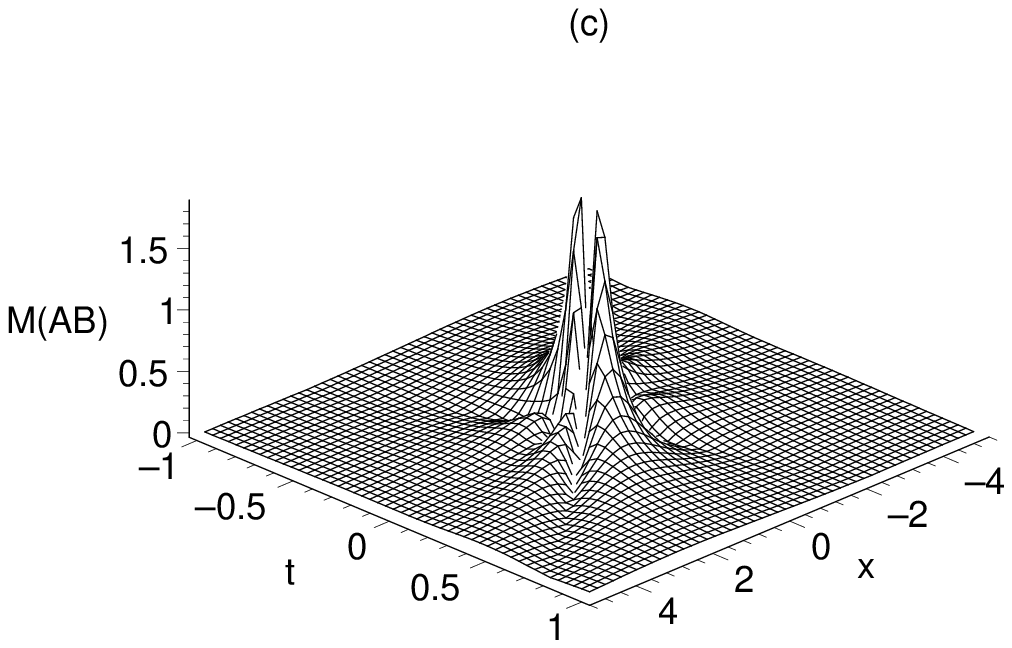}
\epsfxsize=7cm\epsfysize=5cm\epsfbox{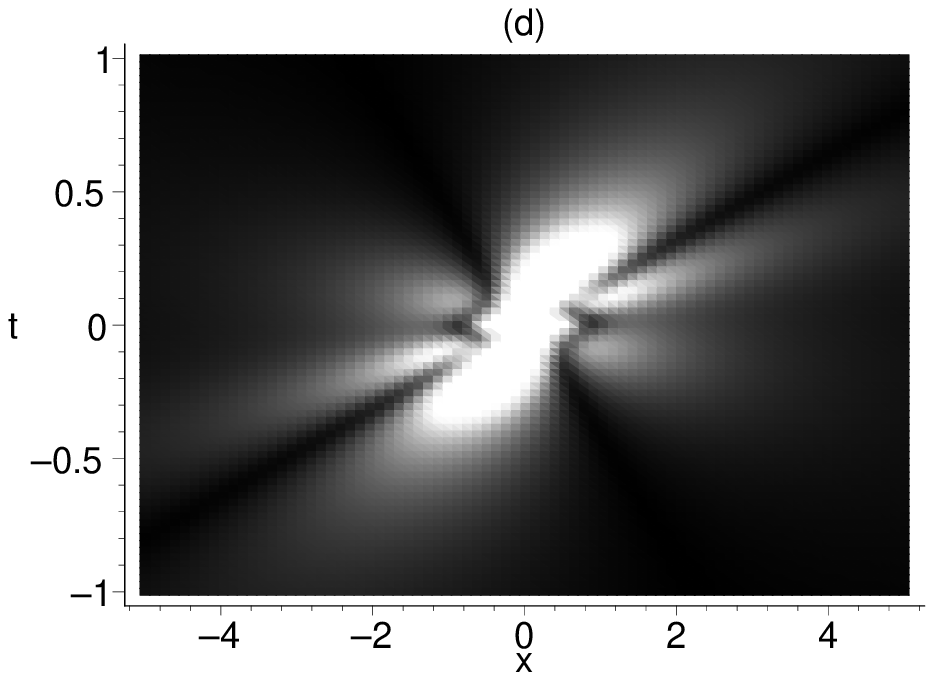}
\caption{Rogue-wave solution  \eqref{RW} with the parameter selections \eqref{ccR}. (a) Real part, (b) imaginary part, (c) amplitude and (d) the density plot of (c).}
\end{figure}

For the general complex parameter selections of $\kappa$ and $\rho$, the solution \eqref{RW} 
may still be an analytic rogue wave or a singular rogue wave dependent on the selections of parameters. Fig. 4 displays the structure for an analytical rogue wave \eqref{RW} with the complex parameter selections.  

\subsection{Two-soliton solutions from zero seed.}

In this section, we  study some kinds of two-soliton solutions. To find two-soliton solutions of the AB-mKdV equation \eqref{ABmKdV}, by means of the 2-fold Darboux transformation,  after solving the constraint condition \eqref{sk}, we have $\gamma_1=-e^{-\frac{1}{2}\xi_1(x_0,t_0)}$, $\gamma_2=e^{-\frac{1}{2}\xi_2(x_0,t_0)}$, $\gamma_3=-e^{-\frac{1}{2}\xi_3(x_0,t_0)}$ and $\gamma_4=e^{-\frac{1}{2}\xi_4(x_0,t_0)}$. Finally, we can obtain the following  soliton-periodic solution
\begin{equation}
A^{[2]}=\frac{-2(\lambda_{23}\lambda_{24}\lambda_{34}e^{\xi_{1}}+\lambda_{13}\lambda_{14}\lambda_{34}e^{\xi_{2}}+\lambda_{12}\lambda_{14}\lambda_{24}e^{\xi_{3}}+\lambda_{12}\lambda_{13}\lambda_{23}e^{\xi_{4}})}{(e^{\xi_{1}+\xi_{3}}+e^{\xi_{2}+\xi_{4}})\lambda_{13}\lambda_{24}+(e^{\xi_{1}+\xi_{2}}+e^{\xi_{3}+\xi_{4}})\lambda_{12}\lambda_{34}+(e^{\xi_{2}+\xi_{3}}+e^{\xi_{1}+\xi_{4}})\lambda_{14}\lambda_{23}},\label{2s}
\end{equation}
where
\begin{eqnarray}
\begin{aligned}
&\xi_{j}=2i\lambda_{j}\left[\left(x-\frac{x_{0}}{2}\right)+4\lambda_{j}^{2}\left(t-\frac{t_{0}}{2}\right)\right],\ \ \  (j=1,2,3,4)\\
&\lambda_{ij}=\lambda_{i}-\lambda_{j}\ \ \   (i,\ j=1,\ 2,\ 3,\ 4\ \ \ and\ \ \ i<j)
\end{aligned}
\end{eqnarray}
It is interesting that the solution \eqref{2s} possesses many interesting properties. Usually it is a singular periodic-solitary wave. For instance, if we select the spectral parameters as 
\begin{equation}
\lambda_1=2,\ \lambda_2=2.1,\ \lambda_3=-\lambda_4=2i,\ x_0=t_0=0, 
\end{equation}
then the solution \eqref{2s} becomes an interaction solution between an soliton and a singular periodic wave as shown in density plot Fig.5 for the quantity 
$$M(A)\equiv \sqrt{\Re(A^{[2]})^2
+\Im(A^{[2]})^2}.$$ 

\input epsf
     \begin{figure}
     \epsfxsize=7cm\epsfysize=5cm\epsfbox{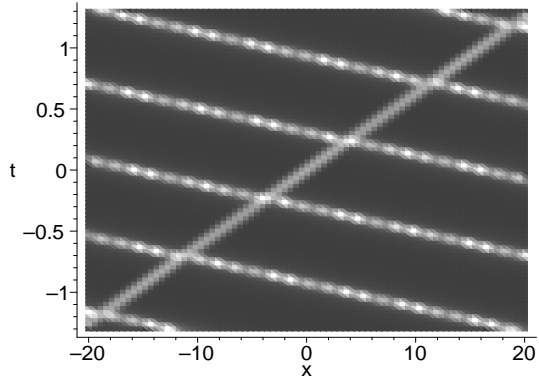}
\caption{The density plot of the interaction between a soliton (analytic) and a singular periodic wave described by  \eqref{2s} with the parameter selections $\lambda_1=2,\ \lambda_2=2.1,\ \lambda_3=2i,\ \lambda_4=-2i,\ x_0=t_0=0$.}
\end{figure}

If  two pairs of spectral parameters are complex conjugate each other, say, $\lambda_3=\lambda_1^*,\ \lambda_4=\lambda_2^*$, then
the solution becomes an analytic  interaction two soliton solution. 
Fig. 6 displays such kind of interaction with the parameter selections 
\begin{equation}
\lambda_1=1+i,\ \lambda_2=1-i,\ \lambda_3=2i,\ \lambda_4=-2i, x_0=t_0=0. \label{cc1}
\end{equation}

\input epsf
     \begin{figure}
     \epsfxsize=7cm\epsfysize=5cm\epsfbox{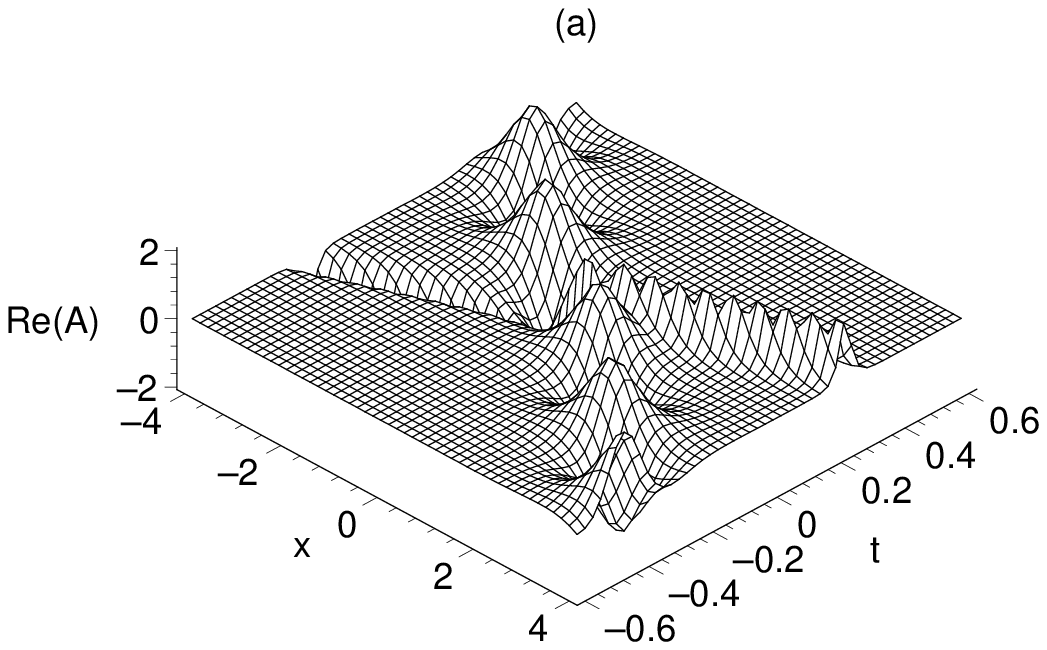}
     \epsfxsize=7cm\epsfysize=5cm\epsfbox{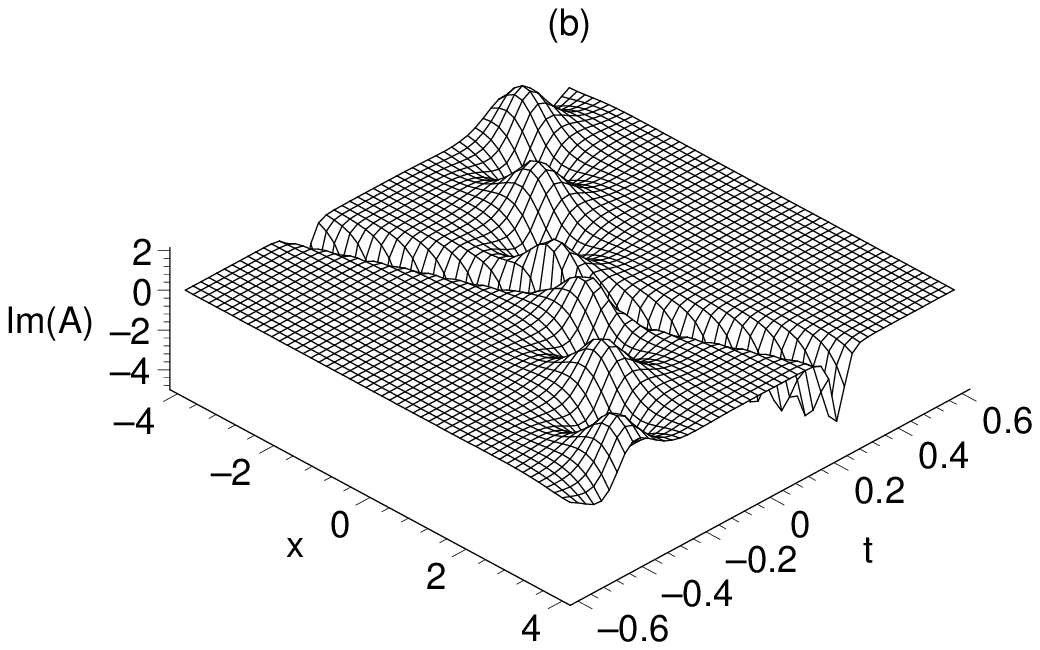}
      \epsfxsize=7cm\epsfysize=5cm\epsfbox{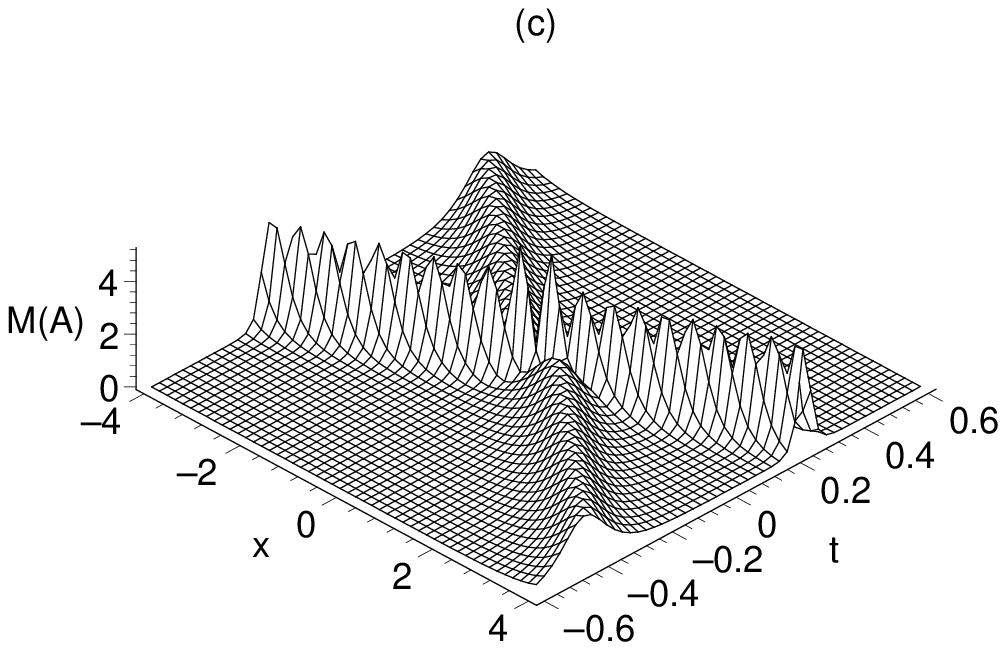} 
\caption{The plot of a standard two soliton interaction solution described by \eqref{2s} with the parameter selections \eqref{cc1}. (a) Real part, (b) imaginary part and (c) amplitude. }
\end{figure}

Fig. 6a, Fig. 6b and Fig. 6c display the structures of \eqref{2s} with \eqref{cc1} for the real part 
$\Re (A^{[2]})$, imaginary part $\Im (A^{[2]})$ 
and the amplitude 
$\left|A^{[2]}\right|
\equiv M(A)$ 
respectively.
From Fig. 6c, we know that this kind of interaction display the usual elastic interaction property. Two solitons have not changed  their velocities and directions of propagation while a phase shift for every soliton will be companied. 

If we select one spectral parameter as zero and the left two spectral parameter are complex conjugate, then the solution \eqref{2s} becomes an interaction solution between a soliton and a kink. 
Fig. 7 and Fig. 8 exhibits two 
different types of interaction modes.

Fig.7 is a plot of \eqref{2s}
with the parameter selection 
\begin{equation}
\lambda_4=x_0=t_0=0, \ \lambda_1=-\lambda_2=i,\ \lambda_3=1.5i. \label{fg7}
\end{equation}

\input epsf
     \begin{figure}
     \epsfxsize=7cm\epsfysize=5cm\epsfbox{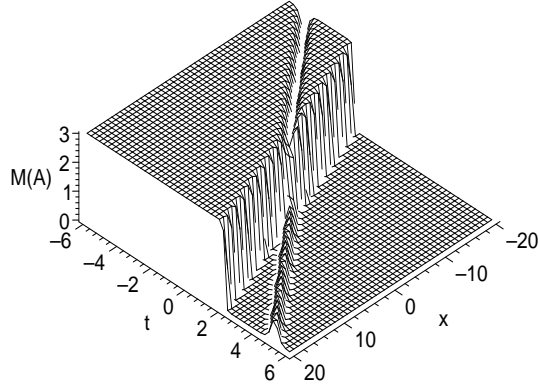}
\caption{The plot of an interaction solution between a soliton and kink described by \eqref{2s} with the parameter selections \eqref{fg7} for the amplitude.}
\end{figure}

From Fig.7, we find that before the interaction we have a dark (gray) soliton and a kink. However, after the interaction, the soliton becomes a bright soliton while the kink remains its shape. This kind of transition comes from the nonlocal interaction of the model. 

Fig.8 is a plot of \eqref{2s}
with the parameter selection 
\begin{equation}
\lambda_2=x_0=t_0=0, \ \lambda_4=-\lambda_3=2i,\ \lambda_1=i. \label{fg8}
\end{equation}

\input epsf
     \begin{figure}
     \epsfxsize=7cm\epsfysize=5cm\epsfbox{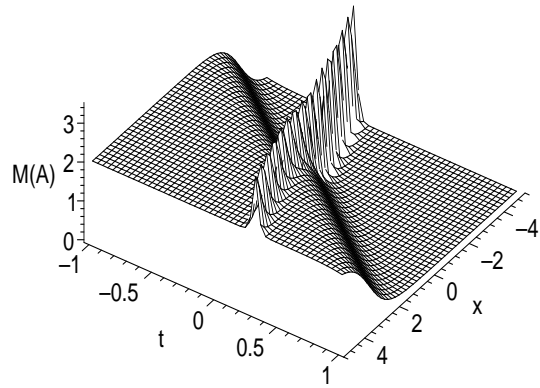}
\caption{The plot of the second type of interaction solution between a soliton and kink described by \eqref{2s} with the parameter selections \eqref{fg8}.}
\end{figure}

From Fig. 8 we can find that the kink is unchanged by the interaction except for the phase shift and the soliton is a bright soliton before and after interaction. However, the soliton is a bright soliton without background before interaction while after the interaction the background of the soliton is not zero because of the existence of the kink.

\section{summary and discussion}
It is shown that the AB systems are important in not only mathematics but also physics. 
In this paper, we  investigate only a special AB-mKdV system which is directly obtained from the third order AKNS system to describe two-place physical events. 
The Nth Darboux transformation of the AB-mKdV system is constructed. 
Some special kind of exact solutions related to the first and second Darboux transformation are explicitly discussed in detail. 
It is found that the complex AB-mKdV equation possesses abundant solution structures which have not yet be noticed before. 
For instance, for single soliton solution of the complex AB-mKdV equation which may be a bell-ring shape soliton (bright soliton) and a kink soliton (dark soliton) dependent on the spectral parameter (but not model  parameters) selections. For the single bell shape soliton, there may be some quite different ones. In this paper two single bell shape bright solitons are obtained. The first one is comes from the first Darboux transformation (Fig. 1) and the second one comes from the second Darboux transformation. 

For the single kink shape soliton, there may be some interesting structures. Especially, the kink solutions of the complex AB-mKdV system may possesses an oscillated tail as shown in Fig. 2c. 

For the rogue wave solutions of the complex AB-mKdV system, its structure is also quite complicated. Two types of structures of the single rogue wave are plotted in the figures 3 and 4 respectively.   
 
For the interactions between two solitons, there are also some interesting phenomena. For instance, the dark soliton may be transition to bright soliton after interaction. 

From the results of this paper, we can see that there are various problems should be revealed in the future studies. 

\section*{Acknowledgements}
The author is grateful  to thank Professors D. J. Zhang, Z. N. Zhu, Q. P. Liu, X. B. Hu, and Y. Chen for their helpful discussions. The work was sponsored by the Global Change Research
Program of China (No.2015CB953904),  Shanghai Knowledge Service Platform for Trustworthy Internet of Things (No. ZF1213), the National Natural Science Foundations of China (No. 11435005) and K. C. Wong Magna Fund in Ningbo University.

\end{document}